\documentclass[12pt]{iopart}
\usepackage{graphicx}

\usepackage{iopams}  
\usepackage{harvard}
\citationmode{abbr}
\begin{document}

\title[Large energy acceptance gantry for proton therapy]{Large energy acceptance gantry for proton therapy utilizing superconducting technology}

\author{K.~P.~Nesteruk, C.~Calzolaio, D.~Meer, V.~Rizzoglio, M.~Seidel, and J.~M.~Schippers}

\address{Paul Scherrer Institut, Villigen PSI, Switzerland}
\ead{konrad.nesteruk@psi.ch}
\vspace{10pt}
\begin{indented}
\item[]December 2018
\end{indented}

\begin{abstract}
When using superconducting (SC) magnets in a gantry for proton therapy, the gantry will benefit from some reduction in size and a large reduction in weight. In this contribution we show an important additional advantage of SC magnets in proton therapy treatments. We present the design of a gantry with a SC bending section and achromatic beam optics with a very large beam momentum acceptance of $\pm15\%$.  Due to the related very large energy acceptance, approximately 70\% of the treatments can be performed without changing the magnetic field for synchronization with energy modulation. In our design this is combined with a 2D lateral scanning system and a fast degrader mounted in the gantry, so that this gantry will be able to perform pencil beam scanning with very rapid energy variations at the patient, allowing a significant reduction of the irradiation time.

We describe the iterative process we have applied to design the magnets and the beam transport, for which we have used different codes. \textit{COSY Infinity} and~\textit{OPAL} have been used to design the beam transport optics and to track the particles in the magnetic fields, which are produced by the magnets designed in~\textit{Opera}. With beam optics calculations we have derived an optimal achromatic beam transport with the large momentum acceptance of the proton pencil beam and we show the agreement with particle tracking calculations in the 3D magnetic field map.

A new cyclotron based facility with this gantry will have a significantly smaller footprint, since one can refrain from the degrader and energy selection system behind the cyclotron. In the treatments, this gantry will enable a very fast proton beam delivery sequence, which may be of advantage for treatments in moving tissue.

\end{abstract}

%
% Uncomment for keywords
\vspace{2pc}
%
% Uncomment for Submitted to journal title message
\submitto{\PMB}
%
% Uncomment if a separate title page is required
%\maketitle
% 
% For two-column output uncomment the next line and choose [10pt] rather than [12pt] in the \documentclass declaration
%\ioptwocol
%

\section{Introduction}
Proton therapy is one of the radiation treatment modalities, which aims for irradiating tumors with a millimeter precision. Charged particles like protons deposit more energy as they slow down, reaching a maximum (the Bragg peak) near the end of their penetration range. In a few millimeters behind the maximum the energy deposition drops very fast to zero and thus the dose is well localized in depth and with a small lateral spread. These features allow the tumors to be irradiated with a maximum efficiency while sparing healthy tissues and surrounding organs at risk.

An accelerated proton beam is transported through a beamline from the accelerator (typically a cyclotron or synchrotron) to the patient. The final section of the beamline is a rotating system called gantry.  The dose prescribed by a physician can be delivered to the tumor volume by different techniques. In modern facilities pencil beam scanning combined with beam energy change is used. In a gantry pencil beam scanning has been first employed for clinical routine at PSI~\cite{pedroni}. In lateral directions the beam is deflected by scanning magnets, which are located before (upstream scanning) or after (downstream scanning) the final bending magnet of the gantry. Pencil beam scanning is usually performed in two lateral dimensions. The location in depth of the Bragg peak is set by choosing the beam energy. In a vast majority of the currently operating proton therapy centers the beam is accelerated either by a cyclotron with fixed extraction energy or a synchrotron with adjustable energy. In cyclotron-based facilities the energy is reduced by a degrader located in the beam transport system, followed by an energy selection system to limit the energy spread in the beam. The latter is is necessary to provide transport of the beam with minimum loses to the patient. 

Gantries for proton therapy are typically at least 100-200 ton devices of more than 10~m in diameter. Installation of such a large and complex system generates various problems and associated costs, such as those related to shielding and building. Therefore, many efforts have recently been taken to reduce both the size and weight of gantries~\cite{gerbershagen2016a,wan}. One of the most promising solutions is the use of superconducting (SC) magnets, which can generate strong magnetic fields keeping the size of coils relatively small. SC magnets do not require water cooling. On the other hand, superconducting magnets have to be cooled by means of a cryogen, such as liquid helium, which introduces some complication to rotation of the gantry. The use of a cryocooler, however, allows SC magnets to be operated without liquid helium. Therefore, according to several groups~\cite{nirs,pronova,oponowicz,trbojevic} the benefits of superconducting technology, such as a smaller footprint and the weight reduced by an order of magnitude, are expected to be dominant.       

Time needed to change the energy is another important parameter of the beam delivery system, which has implications for the whole treatment process. For instance, fast energy changes allow volumetric rescanning of moving targets to be applied in an efficient way~\cite{klimpki}. Other motion mitigation techniques, such as breath-hold or gating, as well as their combinations, also become more efficient with increased treatment delivery speed~\cite{GORGISYAN20171121,gating}.

In this paper we present a design of a new gantry which is conceived to address the need for not only reduced weight and size but also shorter treatment times. The first concept of such a gantry was reported by~\citeasnoun{gerbershagen2016b}. In that work they discussed technical and beam optics requirements for a future design and presented some preliminary first order calculations proving feasibility of the project. The current paper reports on a specific design that has been chosen based on the former considerations. A particular focus is given to our methodology of higher order beam optics design and the results of its optimization. We also present some outlook for realization of the ultra fast energy degrading and scanning systems to complete the concept of the whole design.

Due to the specific beam optics, the gantry is characterized by a large energy (or momentum) acceptance. This feature allows the elimination of a change of magnet setting in the gantry with beam energy during the treatment. Since such magnetic field changes are the major limiting factor determining the speed with which the energy is changed, large energy acceptance enables ultra fast energy setting. Also one avoids many complications and extra cooling power associated with fast field changes of SC magnets. In the next sections the main features of the gantry are presented in more detail, as well as the dedicated design of its beam optics.

\section{Features of the new SC gantry}

\begin{figure}[h]
\centering
\includegraphics[width=0.8\textwidth]{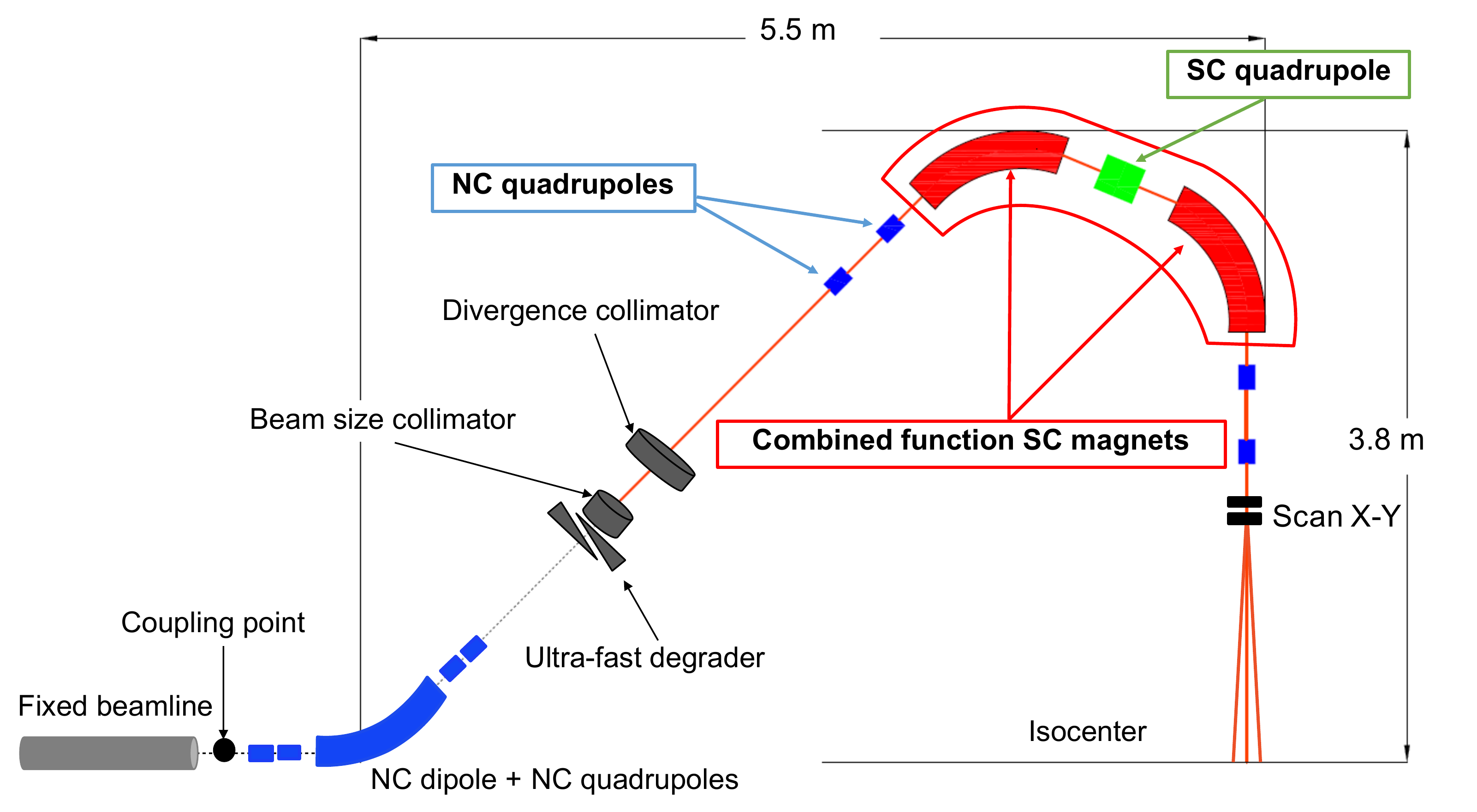} 
\caption{Scheme of a new superconducting gantry with a large momentum acceptance.}
\label{fig:gantry_scheme}
\end{figure}

The new gantry (figure~\ref{fig:gantry_scheme}) consists of a superconducting section, two quadrupole doublets of normal-conducting (NC) magnets, scanning magnets, a beam collimation system, and an integrated ultra-fast energy degrader.
At the degrader location the beam is focused, which allows refraining from a separate energy selection system so that the footprint of the facility can be reduced. In our gantry such an energy selection to limit the energy spread is not needed due to the large energy acceptance. Before the beam reaches the degrader, it is bent by approximately 45 degrees by means of a NC dipole in between two quadrupole doublets. Since the beam at the gantry coupling point has always the same energy of about 230 or 250~MeV with a small energy spread directly from the accelerator, a permanent magnet dipole with a relatively small aperture is sufficient. For synchrotrons, a standard dipole magnet would have to be used due to the variable beam energy. In that case, the degrader could be thinner resulting in reduced scattering loses. The footprint of a facility with such a gantry would be smaller than most of existing solutions. In fact, the higher field strength attainable with superconducting magnets has only a small effect on the radial dimension of the gantry, since it only reduces the bending radius of the proton beam. A reduction of the bending radius from approximately 1~m to about 0.5~m would result in a decrease of the total gantry radius by only 0.5~m, as already mentioned by~\citeasnoun{gerbershagen2016b}.  Another reason is that our large energy acceptance beam optics required a scanning system downstream of the SC bending section, which needs a not too small source-to-axis distance (SAD). Careful considerations, cost estimates and simulations of the effects of the associated large deflection angles in the scanning system have to be performed to determine the optimal combination of the minimum reasonable SAD and field size. The radius and length of the proposed gantry are at the moment estimated to be 3.8~m and 5.5~m, respectively, as shown in figure~\ref{fig:gantry_scheme}. For comparison, Gantry 2 at PSI has a radius of 3.2~m and is 8.9~m long~\cite{gantry2}. Gantry 2, however, is equipped with an upstream scanning system, obtaining parallel beams (infinite SAD) and thus reducing the radius. When comparing with two modern commercial gantries, the upstream scanning Proteus One gantry by IBA has a radius of 3.8~m~\cite{proteus}, while the downstream scanning ProBeam gantry by Varian has a radius of about 5~m~\cite{probeam}.  

The weight of the new gantry would be reduced by up to a factor of 10 with respect to proton-therapy gantries utilizing normal-conducting technology, as for instance the PSI Gantry 2 ($\sim 200$ tons). Among many advantages, the low weight of the magnet allows a simpler, less rigid support structure. Although at this stage it is difficult to precisely estimate the final weight, a good benchmark is the ProNova SC360 superconducting gantry, which weighs about 25 tons~\cite{pronova}.
    
The core of our gantry design is the superconducting section, which allows the large momentum acceptance to be achieved. It is constituted by two combined function dipole-quadrupole-sextupole (DQS) magnets and one combined function quadrupole-sextupole (QS) magnet in between. Each combined function magnet is based on two superconducting racetrack coils~\cite{calzolaio}. The superconductor to be used for winding is Nb$_{3}$Sn. The magnet working temperature will be about 4.5~K. The possibility of using a cryogen-free system relying on cryocoolers has been investigated and the results of simulations are promising~\cite{calzolaio_preprint}.    

The use of superconductivity allows strong focusing components to be obtained in a large volume of the magnets. By efficient suppressing of the dispersion, an unprecedentedly large momentum acceptance of approximately $\pm 15\%$ can be achieved in a relatively simple system consisting of only two bending magnets. The ProNova SC360 SC gantry has demonstrated an increase in momentum acceptance by utilizing superconductivity~\cite{pronova}. \citeasnoun{trbojevic} also presented a design which has this property. Such a unique feature for proton-therapy gantries is the key aspect of the design proposed by us. As already mentioned, the large momentum acceptance (corresponding to about $\pm 30\%$ in the energy domain) allows magnet rampings to be almost completely eliminated, and thus shorten treatment time drastically.

\subsection{Effect of large energy acceptance on treatments}
\begin{figure}
\centering
\includegraphics[width=0.8\textwidth]{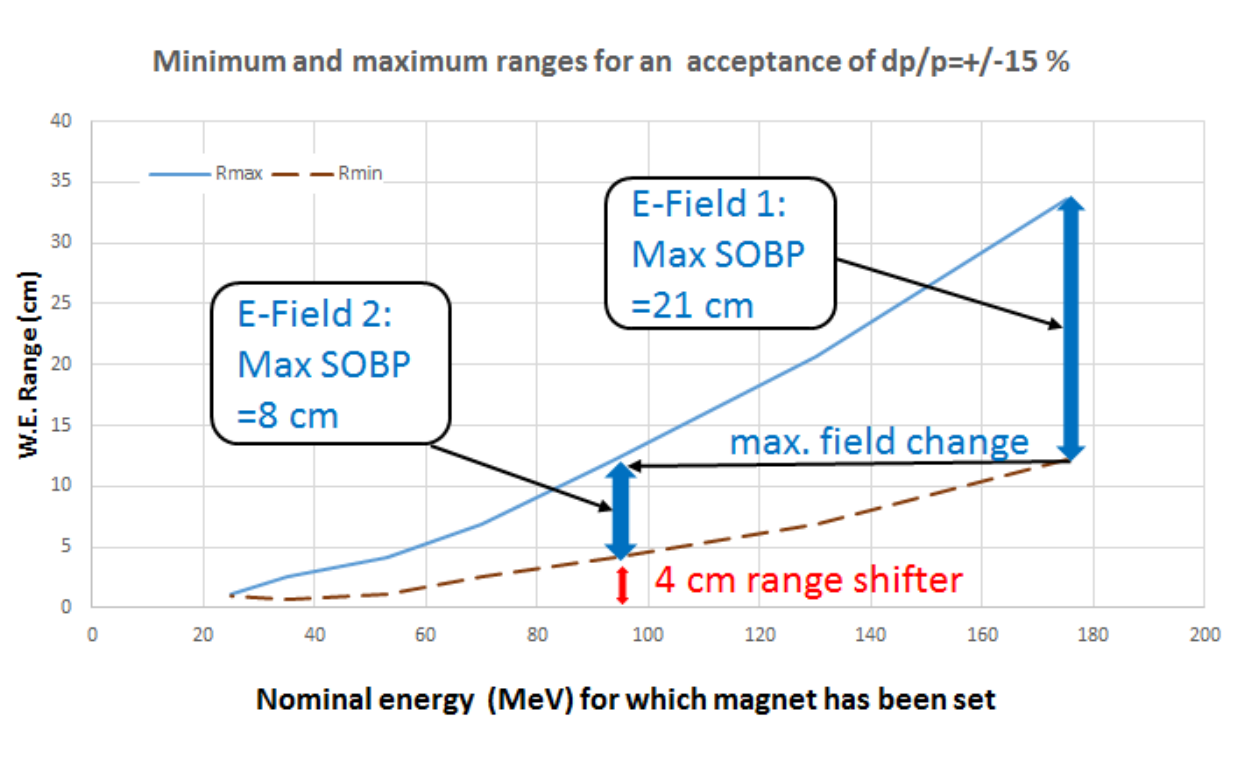} 
\caption{Ranges covered with the degrader as a function of the SC-magnet setting (E-field), assuming a momentum acceptance of $dp/p=\pm15\%$. E-Field~1 is set at the maximum possible SOBP width, covering the ranges 12 and 33~cm at a SC-magnet setting corresponding to a nominal beam energy of 175~MeV. E-Field~2 covers ranges between 4 and 12~cm, if the SC-magnet has been set for the nominal energy of 94~MeV. Depths smaller than the minimum range of 4~cm in E-Field~2 can be covered with a range shifter in the nozzle. It must have a maximum thickness of 4~cm. }
\label{fig:ranges}
\end{figure}

In proton therapy energy modulation is typically done in steps that correspond to a 5 mm change in water equivalent range. In the usual gantries, this also needs a change in the magnetic field strength of approximately 1\%. Due to gantry-magnet design and power supply limitations, such a step can take between 0.1 s (e.g.  at Gantry 2 in PSI) and a few seconds. In the here proposed gantry with the large energy acceptance almost no magnetic field changes are needed. Therefore, the time to change energy is determined by the mechanical speed of the degrader that is mounted on the gantry. Since the incoming (230 or 250 MeV) beam will be focused at the degrader, the beam diameter will be only a few mm, allowing very small lateral degrader movements to make an energy step within a few milliseconds. The momentum acceptance normally used in proton gantries is made such that it can accept the typical momentum spread of 0.5-1\% 2$\sigma$) in a degraded proton beam. The much larger momentum acceptance of the here proposed gantry, however, enables transport of different beam momenta set by the degrader including the associated momentum spread. From the approximate relation between proton range in water and proton beam momentum $dR/R=3.3$~$dp/p$, one can derive that the maximum accepted change in beam momentum ($dp/p=\pm15\%$) covers a variation in w.e. proton range of almost $dR/R=\pm50\%$ with respect to the nominal momentum for which the gantry magnets have been set. For example, in a gantry magnet setting for a nominal energy of 175~MeV, energies between 130~MeV and 230~MeV are accepted. These cover w.e. proton ranges between 12~cm and 33~cm. In figure~\ref{fig:ranges} the thus obtained minimum and maximum ranges have been plotted as a function of the nominal energy for which the gantry magnets are set. 
Depending on the indications selected for treatment at this gantry, the number of magnet settings is only one in most cases. For a treatment needing two magnet settings, one can consider the change between the two magnet settings (''E-fields'') similar as a change of field (gantry angle). For a magnetic field change with 0.1~T/s, which may need some additional cooling power~\cite{calzolaio_preprint}, this takes a similar time as a gantry rotation, so in the order of 10~seconds. The need for covering a limited (max. ~4~cm) extra range can also be satisfied by inserting a range shifter in the nozzle just before the patient. The example in figure~\ref{fig:ranges} shows that a target extending over a depth 4 and 33~cm is to be treated by two magnet settings (''E-fields''). For smaller depths one can insert a range shifter in the nozzle.  

To investigate how many fields are to be used for typical indications for proton therapy, we have used data published by~\citeasnoun{suzuki}, which show the needed SOBP width (i.e. max. variation of range) as a function of the maximum range for four groups of indications. Combining these data with the ranges our design covers by one or two E-fields (from figure~\ref{fig:ranges}), indicates that, for the indications studied, approximately 70\% of the patients can be treated with a single magnet setting. From the published data one can also derive that an eventually necessary magnetic field change to a second E-field is typically in the order of 20\%. This is expected to be possible within 30 seconds, similar as when a gantry angle is changed.  

\subsection{Fast degrader system}

In systems based on a cyclotron, a degrader system is required to modulate the beam energy. This consists of additional material in the beam path with variable thickness, which can be changed continuously or in discrete steps. In addition to the beam energy, the degrader system also effects the beam quality since multiple Coulomb scattering primarily increases the phase space of the beam. Therefore, the location of the degrader system along the beam line is highly critical for the system performance. In addition, the choice of degrader material is also an important system parameter~\cite{boron_carbide}. If a degrader is used without subsequent collimation, the system should be placed as close as possible to the patient with minimal air gap to minimize the effect of the increased beam. In this case, however, the degrader must cover at least the entire scan area and becomes large and heavy. In practice, a small air gap can hardly be realized due to the increased nozzle and the patient geometry.
 
\begin{figure}
\centering
\includegraphics[width=0.8\textwidth]{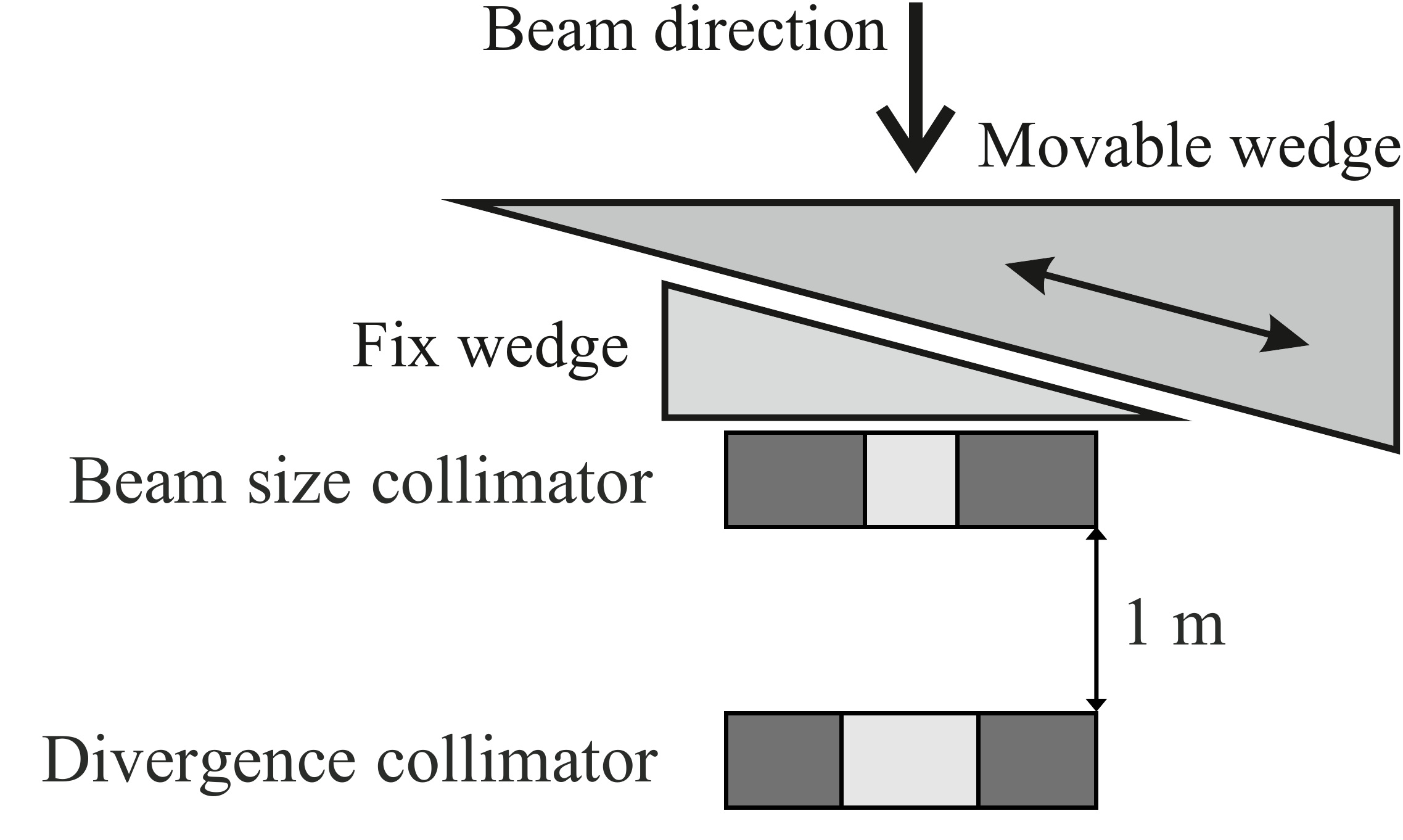} 
\caption{Scheme of the fast degrader system.}
\label{fig:degrader}
\end{figure}

With the proposed gantry optics, the degrader unit can be placed before the superconducting bend and take fully advantage of the large momentum acceptance. The system consists of the degrader itself and two collimators that define the phase space, as shown in figure~\ref{fig:gantry_scheme}. The beam optics provides a 1:1 image from the first collimator to the isocenter. Since the degrader is located in a waist of the beam envelope, followed by a collimator with a diameter $<1$~cm, it is sufficient if the degrader covers a cross-sectional area of less than $2\times2$~cm$^{2}$. Technically such a degrader can be realized as a rotating wheel with increasing material thickness or as a sliding wedge. A second fix wedge, as shown in figure~\ref{fig:degrader}, provides uniform energy losses. The moving mass of such a wedge will be in the order of 1~kg depending on the wedge geometry. Commercially available linear motors are best suited for fast and precise longitudinal motions and will allow to change beam energy in typical steps of 1\% in less than 10 ms.

Beam energy reduction in the degrader as well as beam losses in the collimators will result in an unwanted production of neutrons due to nuclear interactions. This could affect the machine - particularly the superconducting coil - and additional neutron dose to the patient must be avoided. In a simulation study~\cite{talanov}, we have shown that there is a non-negligible increase in the neutron dose at the isocenter, which, could be reduced to a lower value with additional shielding. 

\section{Beam optics design}
In this paper we present our study of the beam optics between degrader exit and isocenter. The design of the beam optics was performed with the \textit{COSY Infinity} simulation software. It is an arbitrary order beam physics code based on differential algebra~\cite{cosy1,cosy2} and allows the computation of beam transport for arbitrarily complicated fields and to arbitrary order. With \textit{COSY Infinity} beam transport matrices of up to the fourth order were calculated and optimized. The corresponding magnetic lattice was modeled in the electromagnetic simulation software~\textit{Opera} by~\citeasnoun{calzolaio_preprint}, using the optimal field shapes and strengths as derived from \textit{COSY Infinity}. It was then optimized in an iterative way by extracting from the Opera model the multipole content (dipole, quadrupole, and sextupole components) along the trajectory of the reference particle and refitting the lattice in~\textit{COSY Infinity}. The solution is validated with the particle tracking software \textit{OPAL}~\cite{opal} which allows particles to be tracked by time integration in field maps generated in \textit{Opera}. The use of different codes allows all the design parameters and calculation results to be kept under better control, which is particularly needed to achieve the large momentum acceptance. It has also given the possibility of performing various consistency checks throughout the whole design process. 

The first feasibility study, reported by~\citeasnoun{gerbershagen2016b}, has been performed with the \textit{Transport} software~\cite{transport,rohrer}. It showed that if the dispersion suppression is already starting within the first bend, it makes the lattice accept a wide energy band. Moreover, it has also indicated that locating the energy degrader at an intermediate image of the gantry coupling point, just before the final bend, would help correct the chromatic imaging errors. These results have been used as the basic concept of the beam optics design. The beam at the beam size collimator, directly following the energy degrader, is imaged to the isocenter. The bending section of our gantry is based on the standard configuration of a bending achromat - dipole-quadrupole-dipole. We have included a stronger dispersion suppression in the bending magnets by adding a horizontally focusing quadropole component to the dipole .

\subsection{Lattice fitting}

The starting point of the simulated optics is the location of the beam size collimator following the energy degarder directly. There the $1\sigma$ transverse beam emittance is assumed to be $2.25~\textrm{mm}\times3.33~\textrm{mrad}=7.5~\textrm{mm~mrad}$ in both horizontal and vertical planes. The nominal energy of the simulated beam is 185~MeV. The simulated 7.8~m long lattice between the starting point and isocenter corresponds to the scheme shown in figure~\ref{fig:gantry_scheme}. The SC part of the lattice is surrounded by two quadrupole doublets arranged symmetrically on either side of the bending section. The drift space between the beam size collimator and the first NC quadrupole is equal to the one between the last NC quadrupole and the isocenter, so that the simulated lattice is completely symmetric with respect to the central SC quadrupole (QS), which plays the key role in getting an chromatic bending with limited higher order aberrations. Within the latter drift space the scanning system will be fitted.

In order to create the first approximate design of the magnetic lattice, a Sharp Cut-Off Fringe Field (SCOFF) model, which thus neglects fields outside the magnets, is simulated in~\textit{COSY Infinity}. In our design we constrained the dispersion at two locations. Of course, as in the case of other existing gantries, the transverse and angular dispersion should be suppressed at the isocenter. Apart from that, we also require the transverse and angular dispersion to be suppressed already after the second bending magnet (DQS~2). This constraints a non-zero dispersion to the SC bending section. It is realized by using relatively strong quadrupole and sextupole components in all the three SC magnets. One of the important constraints is to keep all the SC magnets focusing in the dispersive X plane. This allows the potential of SC magnets to be maximally exploited for the reduction of the geometrical dispersion. The lattice fitting encompasses the following criteria:
\begin{itemize}
\item dispersion suppression at two locations
\item beam waist in the middle of the central quadrupole in the non-dispersive Y plane
\item beam imaging between the collimator and isocenter with a magnification close to unity
\item minimization of higher order components in the beam transport
\end{itemize}

\begin{figure}[p]
\centering
\includegraphics[width=0.95\textwidth]{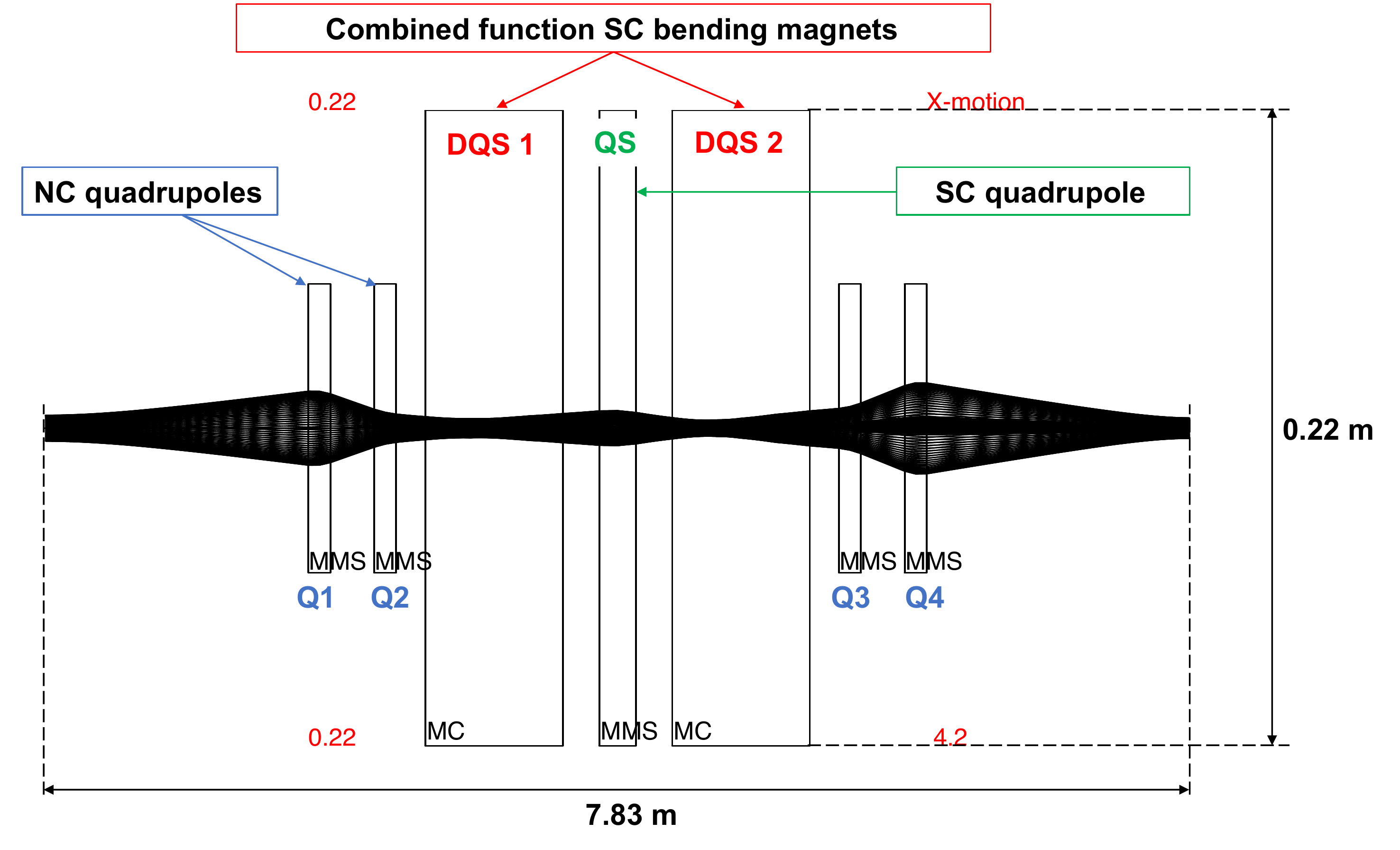} 
\includegraphics[width=0.95\textwidth]{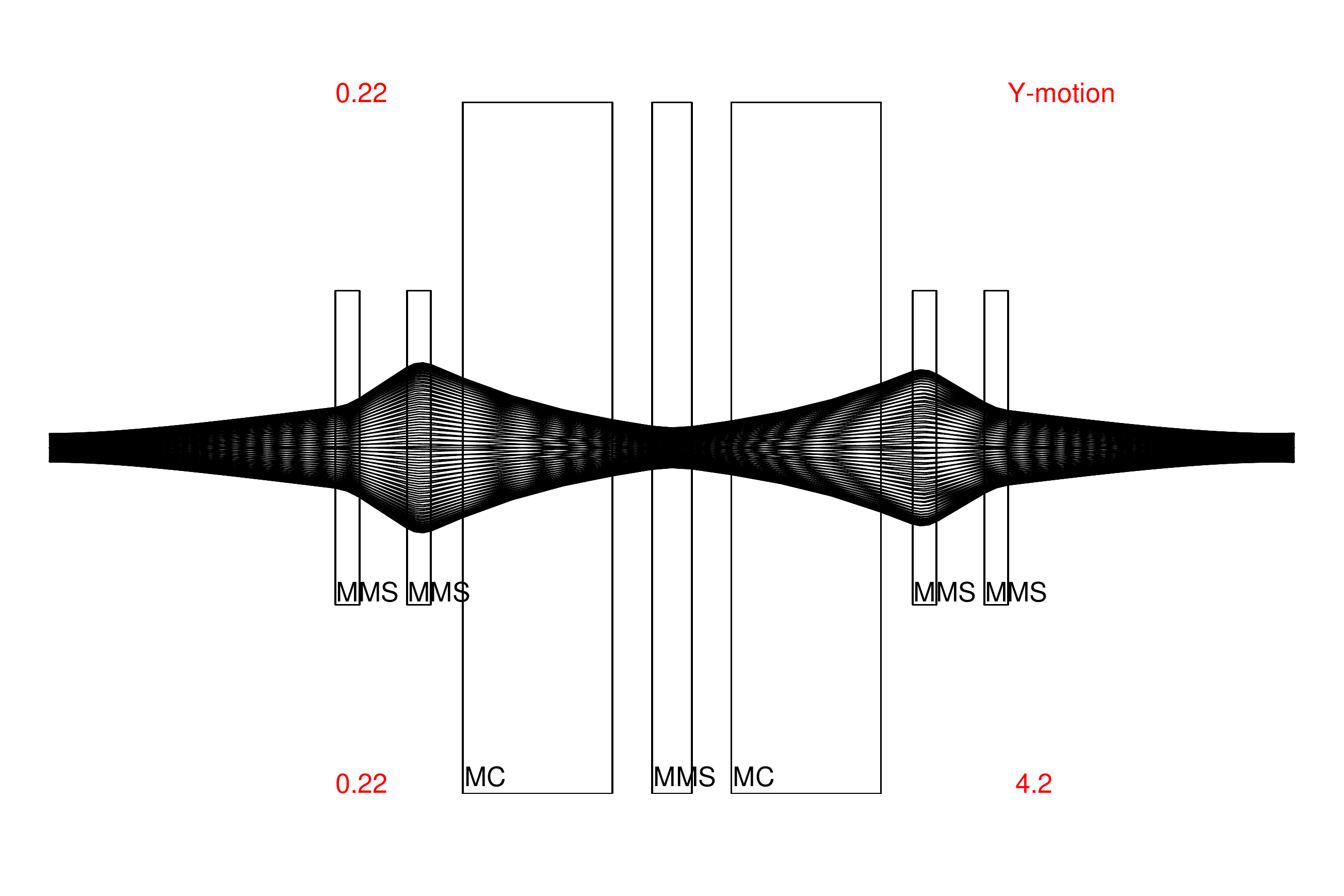}
\caption{Result of the SCOFF model lattice fitting for the nominal momentum in the dispersive X plane (top) and non-dispersive Y plane (bottom). The 2$\sigma$ beam envelope is shown. In the top figure, the lattice components and dimensions are highlighted on the top of the standard~\textit{COSY Infinity} graphical output.}
\label{fig:0scoff}
\end{figure}

\begin{figure}[p]
\centering
\includegraphics[width=0.95\textwidth]{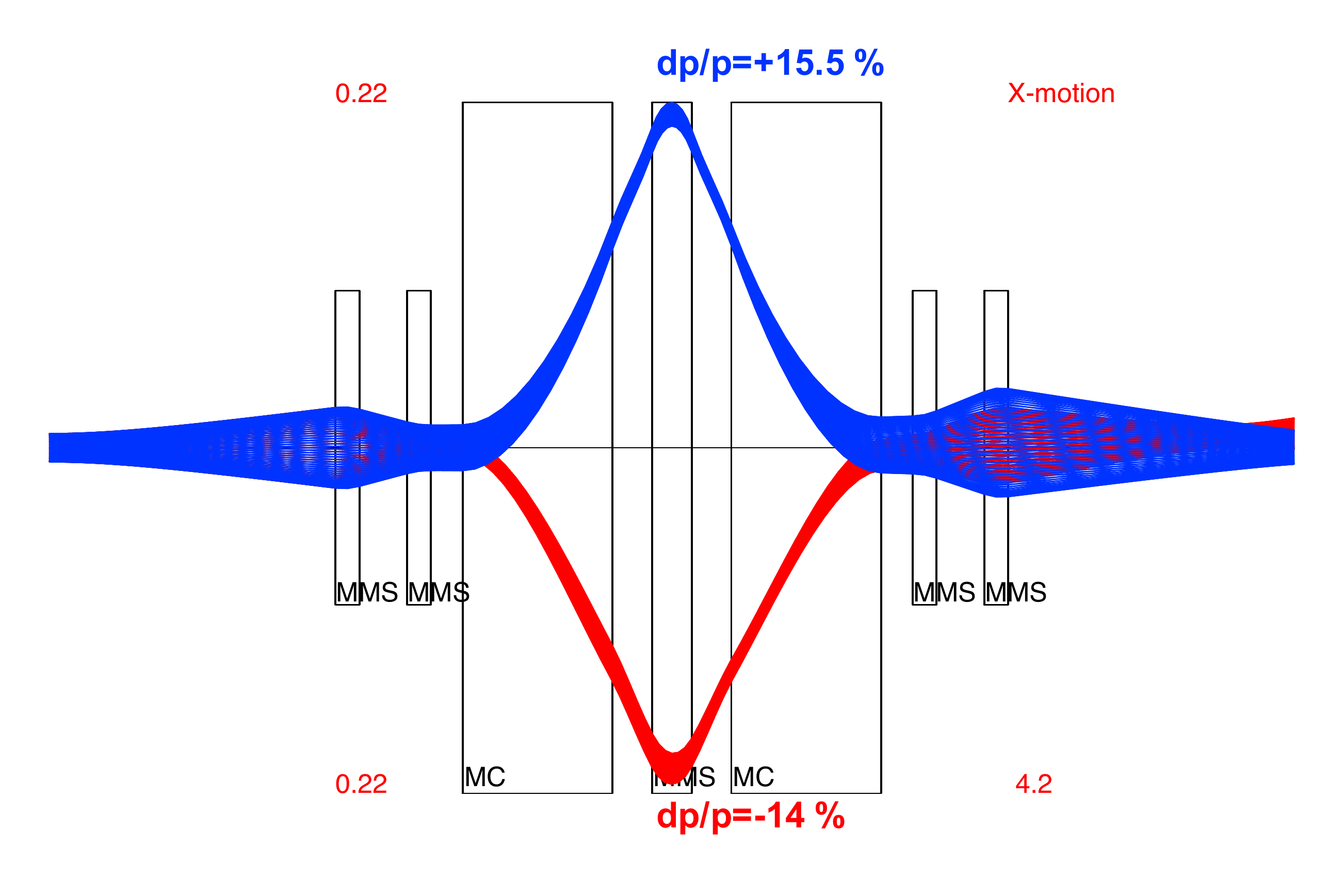} 
\includegraphics[width=0.95\textwidth]{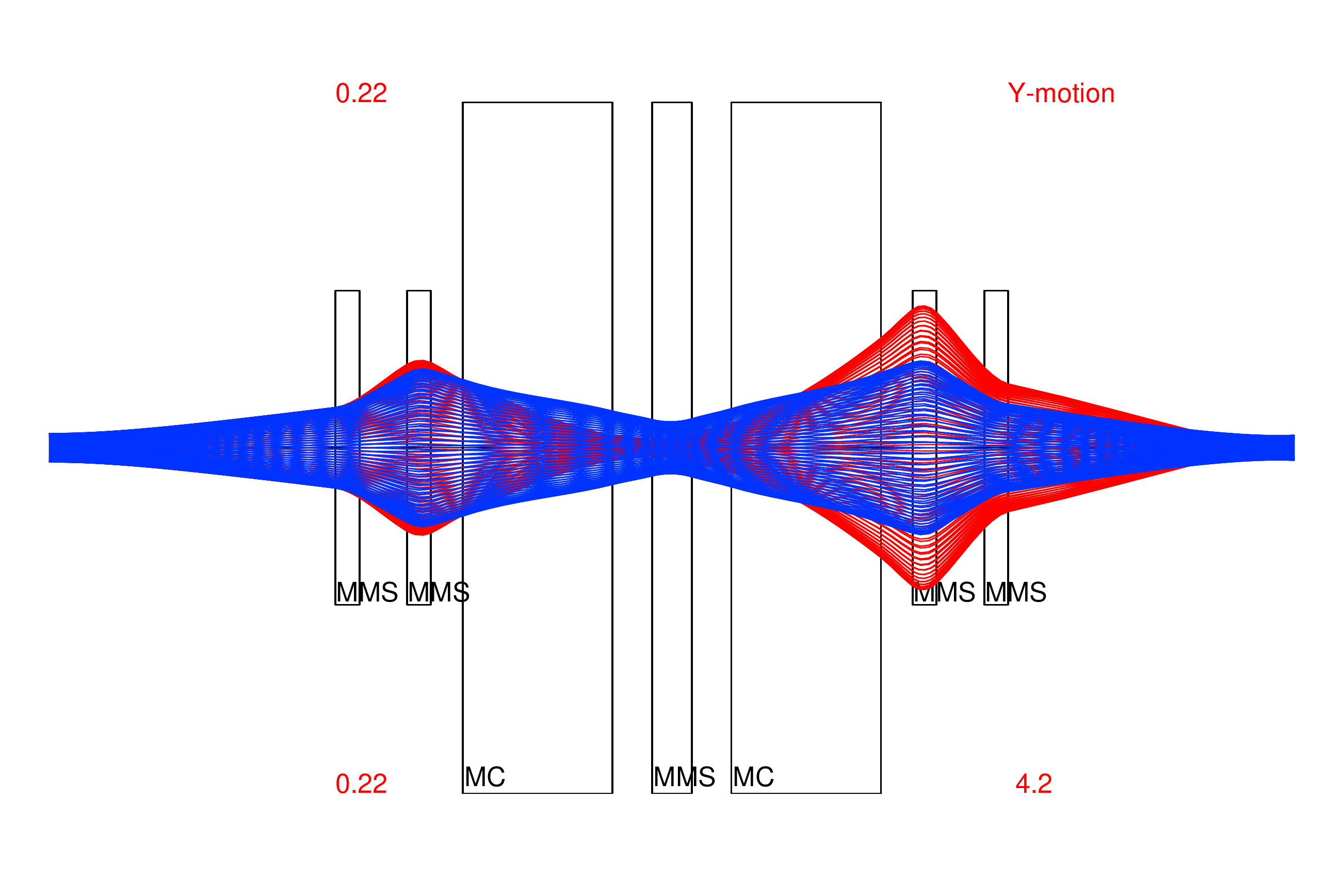}
\caption{Result of the SCOFF model lattice fitting for $dp/p=+15.5\%$ (blue) and $dp/p=-14\%$ (red) in the dispersive X plane (top) and non-dispersive Y plane (bottom). The 2$\sigma$ beam envelope is shown.}
\label{fig:pm15scoff}
\end{figure}

\begin{figure}
\centering
\includegraphics[width=\textwidth]{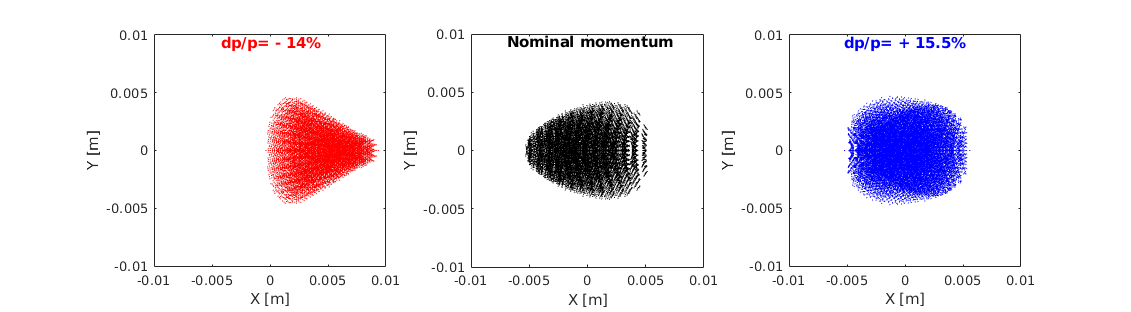} 
\caption{Beam spots at the isocenter for different beam momenta.}
\label{fig:beam_spots}
\end{figure}

The results of the lattice fitting for a SCOFF model and the nominal momentum, with zero momentum spread, are presented in figure~\ref{fig:0scoff}. The 2$\sigma$ beam envelope (beam radius of 4.5~mm) is shown. The dispersion is suppressed to the level of 0.1~mm/\% after DQS 2, as well as at the isocenter. For the chosen aperture sizes, which are considered good magnetic field regions, it corresponds to a momentum acceptance of -14\% and +15.5\%. Figure~\ref{fig:pm15scoff} shows the 2$\sigma$ beam envelope for the momentum deviations $dp/p$ corresponding to the lower and upper limits of the obtained acceptance. Beam spots obtained at the isocenter are shown in figure~\ref{fig:beam_spots}. The visible distortions are mostly due to the sextupole component of the magnetic field, which results in a horizontal (X plane) force acting on the beam with a magnitude that is a quadratic function of the distance from the midplane. The 2$\sigma$ beam sizes are found to be 4.7-5.4~mm and about 3.7-4.0~mm in the X and Y planes, respectively. The maximum variation of the beam sizes does not exceed 13\%, which fulfills the treatment planning requirements. For the highest accepted momentum deviation, an offset (beam displacement) in the X-direction of about 4.7~mm is observed with respect to the nominal momentum beam spot. However, knowing the corresponding energy, such an offset can be easily corrected by the scanning magnets.    


\newpage
\subsection{Magnetic design}
\begin{figure}
\centering
\includegraphics[width=0.8\textwidth]{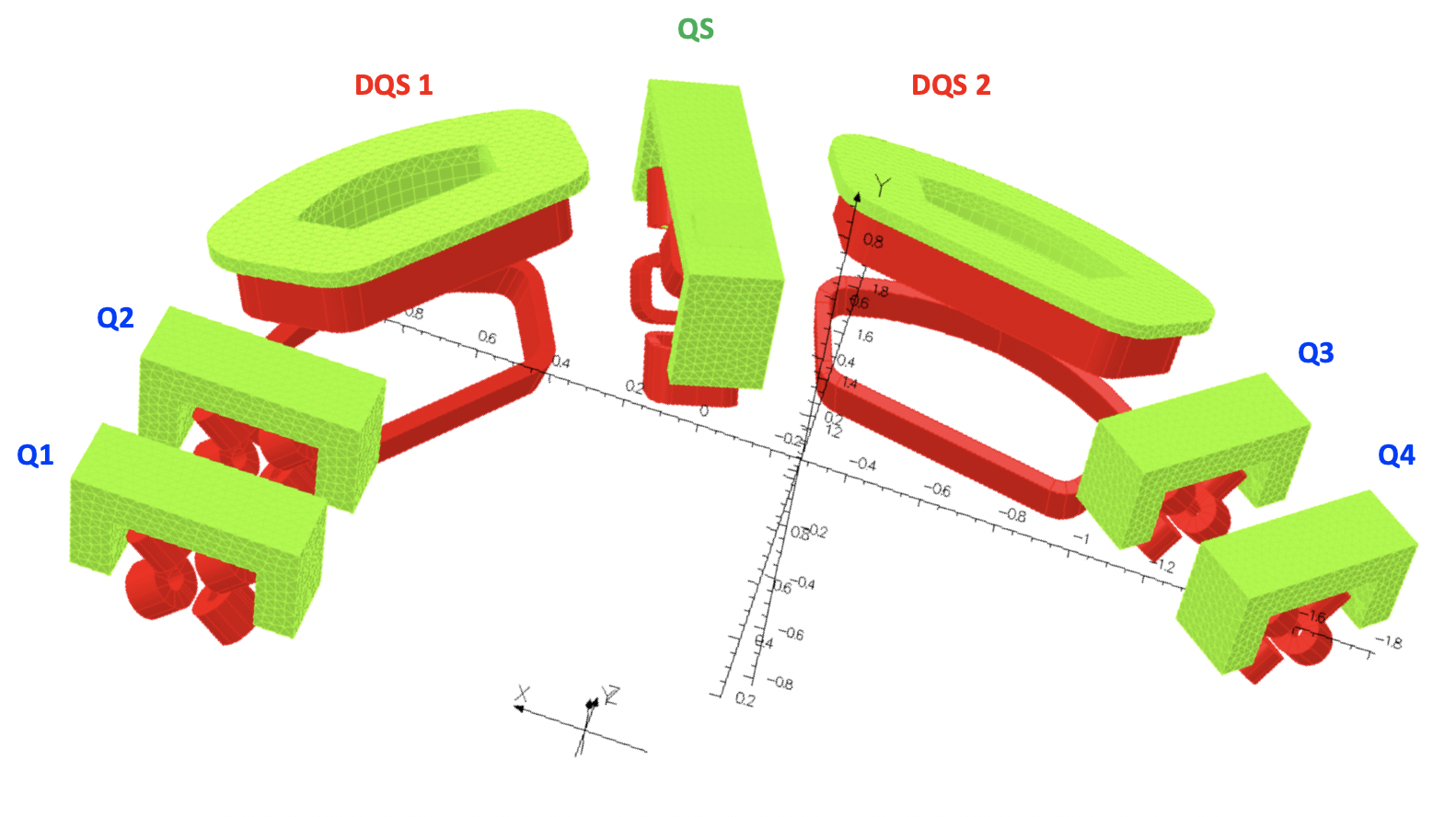} 
\caption{Model of the magnetic lattice designed in\textit{Opera}.}
\label{fig:opera}
\end{figure}

\begin{figure}
\centering
\includegraphics[width=\textwidth]{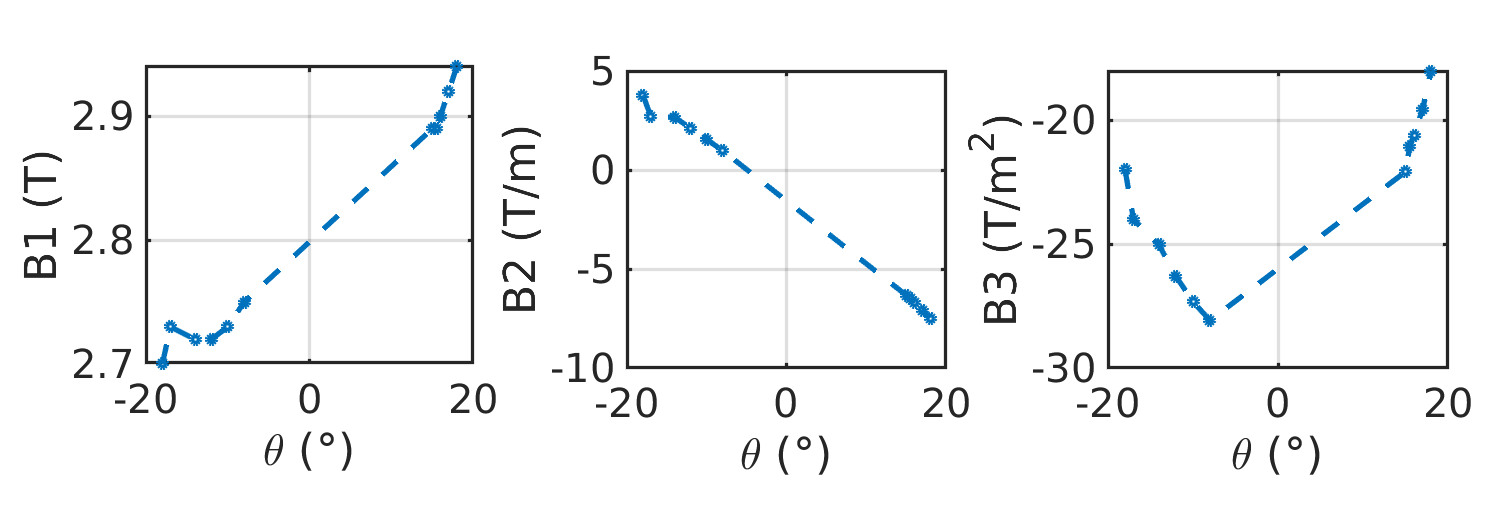} 
\caption{$B1, B2, B3$ as function of the tilt angle $\theta$~\cite{calzolaio_preprint}.}
\label{fig:tilt}
\end{figure}

The multipole values obtained in the SCOFF lattice fitting are used to specify the corresponding magnets in the~\textit{OPERA} software. The design process of the magnetic lattice is described in detail by~\citeasnoun{calzolaio_preprint}. A resulting model is shown in figure~\ref{fig:opera}. 
The most complex are the DQS magnets. They have to produce a field with specified dipole ($B1[\textrm{T}]$), quadrupole ($B2[\textrm{T}/\textrm{m}]$), and sextupole ($B3[\textrm{T}/\textrm{m}^{2}]$) components. The main parameters used to produce the required field harmonics are the horizontal aperture and the tilt angle between the coils. The latter allows the desired quadrupole component to be obtained. The sensitivity of the three components $B1, B2, B3$ to the variation of the tilt angle $\theta$ is shown in figure~\ref{fig:tilt}. In order to introduce focusing in the X plane by the DQS magnets, the tilt angle must be negative so that the field is maximum in the outer region of DQS. As already mentioned, focusing in the X plane widens the momentum acceptance band. 

\begin{figure}
\centering
\includegraphics[width=0.8\textwidth]{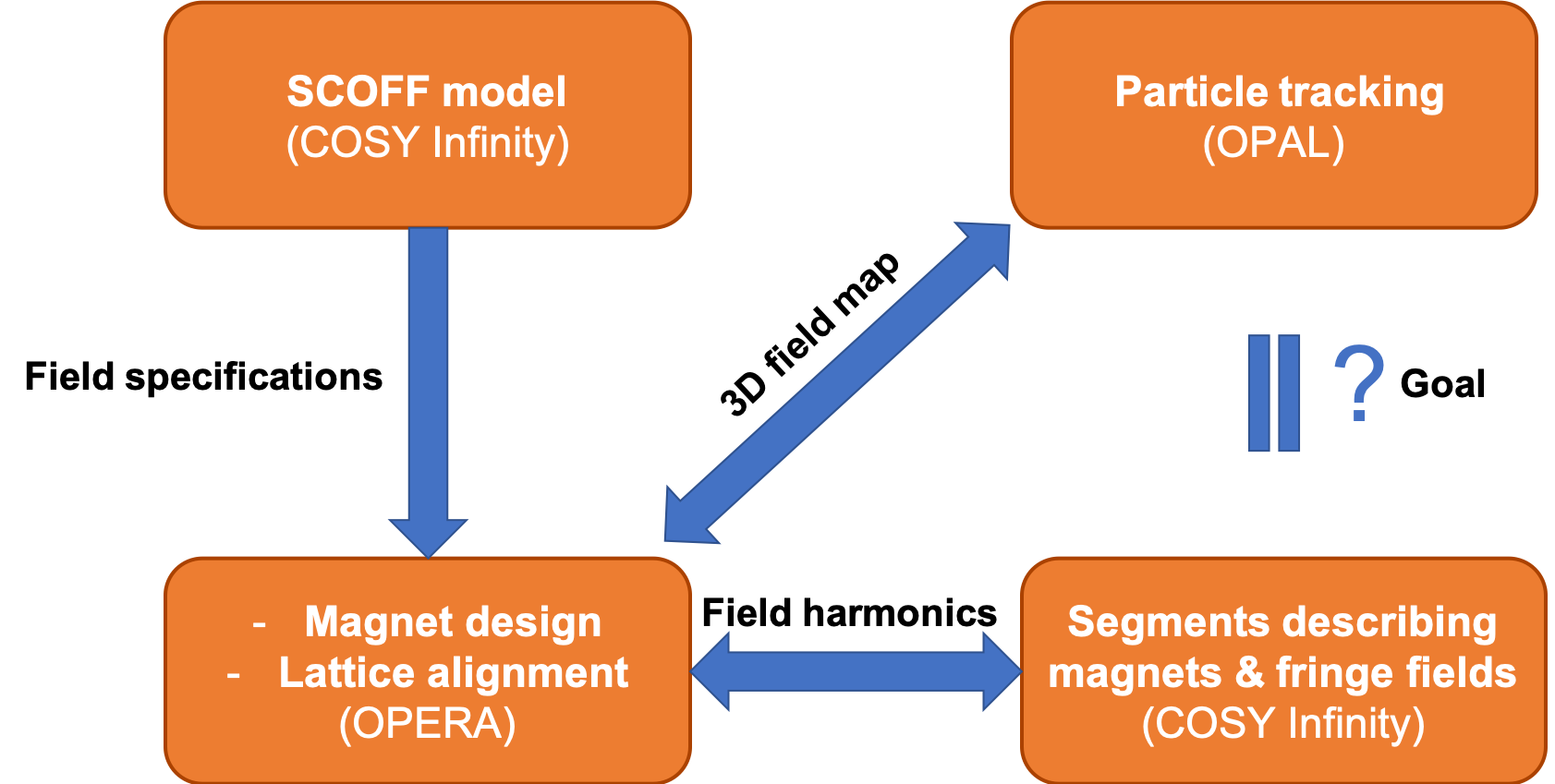} 
\caption{Block diagram of the beam optics design.}
\label{fig:design}
\end{figure}

Given that all the single magnets are optimized according to the input parameters from~\textit{COSY Infinity}, the next essential issue is their alignment. The position and orientation of each magnet is defined in~\textit{Opera} and determines the outcome of simulations in~\textit{COSY Infinity} and~\textit{OPAL} in our design process with different codes, which is depicted in figure~\ref{fig:design}. 
Our lattice is very sensitive to misalignments mostly due to the presence of relatively strong stray fields and due to the fact that the quadrupole magnetic axis in the DQS magnets does not have a constant curvature radius. The former can be reduced by adding more iron. Suppressing the latter requires more complex design of coils so that not only is the tilt angle changed but also their shape. The alignment procedure is described by~\citeasnoun{calzolaio_preprint} and can be summarized in the following steps: 
\begin{enumerate}
\item The magnets are arranged with respect to the reference proton traveling along the magnets' geometrical axis. At this step no field in the drift spaces is considered.
\item Only the DQS magnets are turned on. By shooting protons of energies in the range $185\textrm{~MeV}\pm10\%$, the relationship between the energy of the incident proton and the exit angle of the proton is found. The corresponding correction of the magnetic field is calculated so that the exit angle is minimized.
\item The QS magnet (being still switched off) is moved radially to make protons travel through the magnetic axis, which is shifted with respect to the geometrical one.    
\item \label{nc_alignment} The positions of the NC quadrupoles are adjusted according to the trajectory of the reference proton at the nominal energy.
\item \label{qs_radial} Since the QS magnetic axis is further displaced due to the DQS stray field, the QS is moved radially in an iterative way based on the tracking of the reference particle. The iterative approach is needed as the total 3D map slightly changes while moving the QS magnet due to its iron yoke.
\item All the magnets are turned on. A reference particle is shot again. The distance between the reference track and the magnetic axes of QS and NC quadrupoles is calculated. If it is larger than 0.5~mm, steps~(\ref{nc_alignment}) and~(\ref{qs_radial}) are repeated.  
\end{enumerate}

Once the magnetic lattice is aligned, the magnetic field harmonics up to the sextupole component are evaluated as a function of the location along the lattice. The harmonics are calculated with respect to the reference track, which due to the stray fields is not equivalent to the geometrical axis. This is important, since in \textit{COSY Infinity} all the magnets are assumed to be perfectly aligned and, unless specified explicitly, their geometrical axis coincides with the reference track. At no stage of our design fringe fields are modeled by parameterization.  Therefore, the harmonics are used to define a lattice model, which represents the effects of fringe fields, to be optimized in~\textit{COSY Infinity}. The lattice is divided in 25~mm wide segments. In each segment three multipole components - dipole, quadrupole, and sextupole are defined. The harmonics from the~\textit{Opera} model are calculated with only one magnet turned on at a time, producing 7 separate field maps. Then the superposition principle is applied and the harmonics are added with corresponding map-specific scaling factors, which at the beginning are all set to one. In a segment each multipole component is the sum of the respective components in the 7 maps, each scaled by the scaling factor. So, in total there are 7~(maps)~$\times$~3~(multipoles)~$=$~21 scaling factors. Therefore only the scaling factors are fitted in~\textit{COSY Infinity} while keeping the same shape of the longitudinal profile of each multipole component. The shape of the quadrupole field profiles changes with the tilt angle of the DQS magnets. Therefore the optimization of the segmented model is done in this iterative way. It has to be noted that the superposition principle introduces an uncertainty due to the non-linearity caused by the presence of iron. The difference between the harmonics for all magnets being turned on and for the superposition was estimated to be generally small but locally could reach even 20\%. This discrepancy is one of the reasons for an additional fine tuning of the parameters, which is needed at the last stage of the current design process. Further steps to optimize will be necessary anyway, when the mechanical designs of the coils are ready.  

\begin{figure}
\centering
\includegraphics[width=0.95\textwidth]{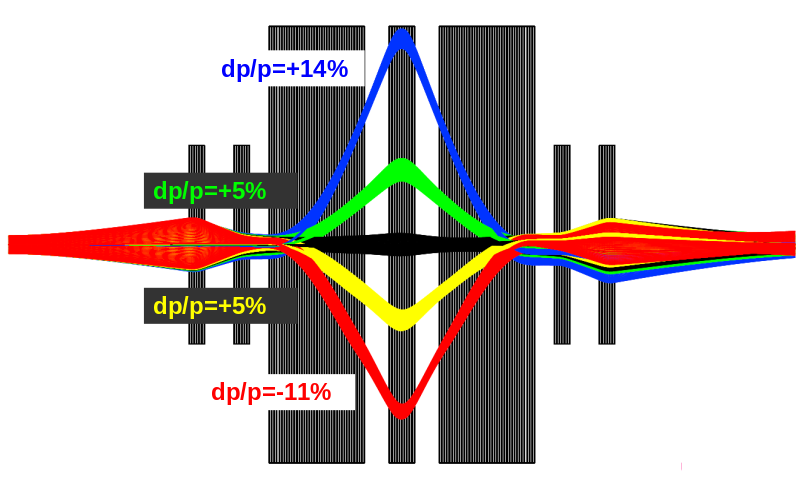} 
\includegraphics[width=0.95\textwidth]{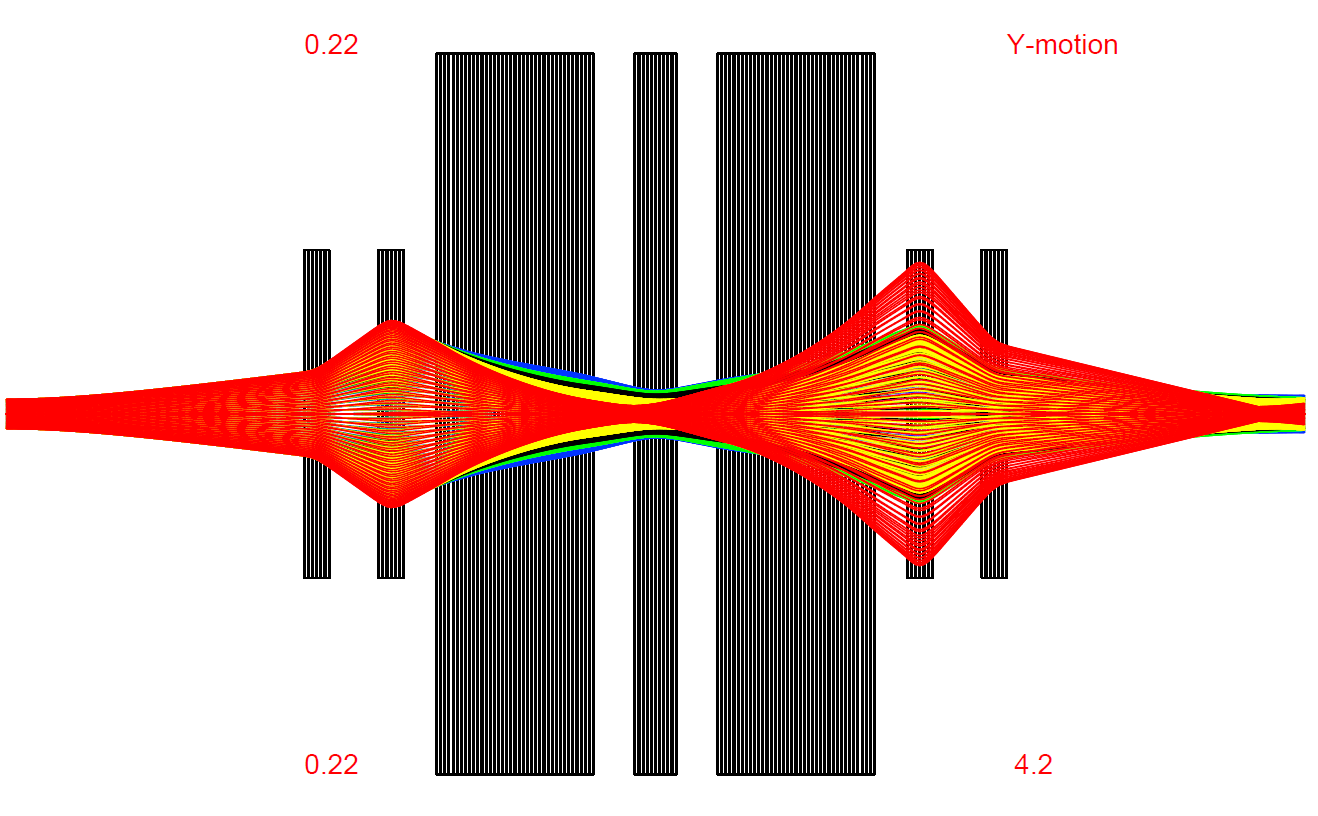}
\caption{Result of the optimization of the segmented model for different momentum deviations in the dispersive X plane (top) and non-dispersive Y plane (bottom). The 2$\sigma$ beam envelope is shown.}
\label{fig:COSYfringe}
\end{figure} 

The results of the optimization of the segmented model in~\textit{COSY Infinity} are shown in figure~\ref{fig:COSYfringe}. The momentum acceptance $dp/p=-11\%-+14\%$ is slightly smaller with respect to the SCOFF model ($dp/p=-14\%-+15.5\%$) with the same good field region and thus the same apertures. As it can be seen figure~\ref{fig:COSYfringe}, for large momentum deviations the beam is not imaged from the starting point to the isocenter, especially in the Y plane. To restore imaging and extend the momentum acceptance beyond $dp/p=-11\%-+14\%$, it is possible to change only the settings of the NC quadrupoles. Such a change of the NC quadrupole settings could easily be implemented without extending the treatment time, since ramping of these NC quadrupoles is very fast and could be done in parallel with changing the energy.

%\newpage
\subsection{Particle tracking}

\begin{figure}
\centering
\includegraphics[width=0.8\textwidth]{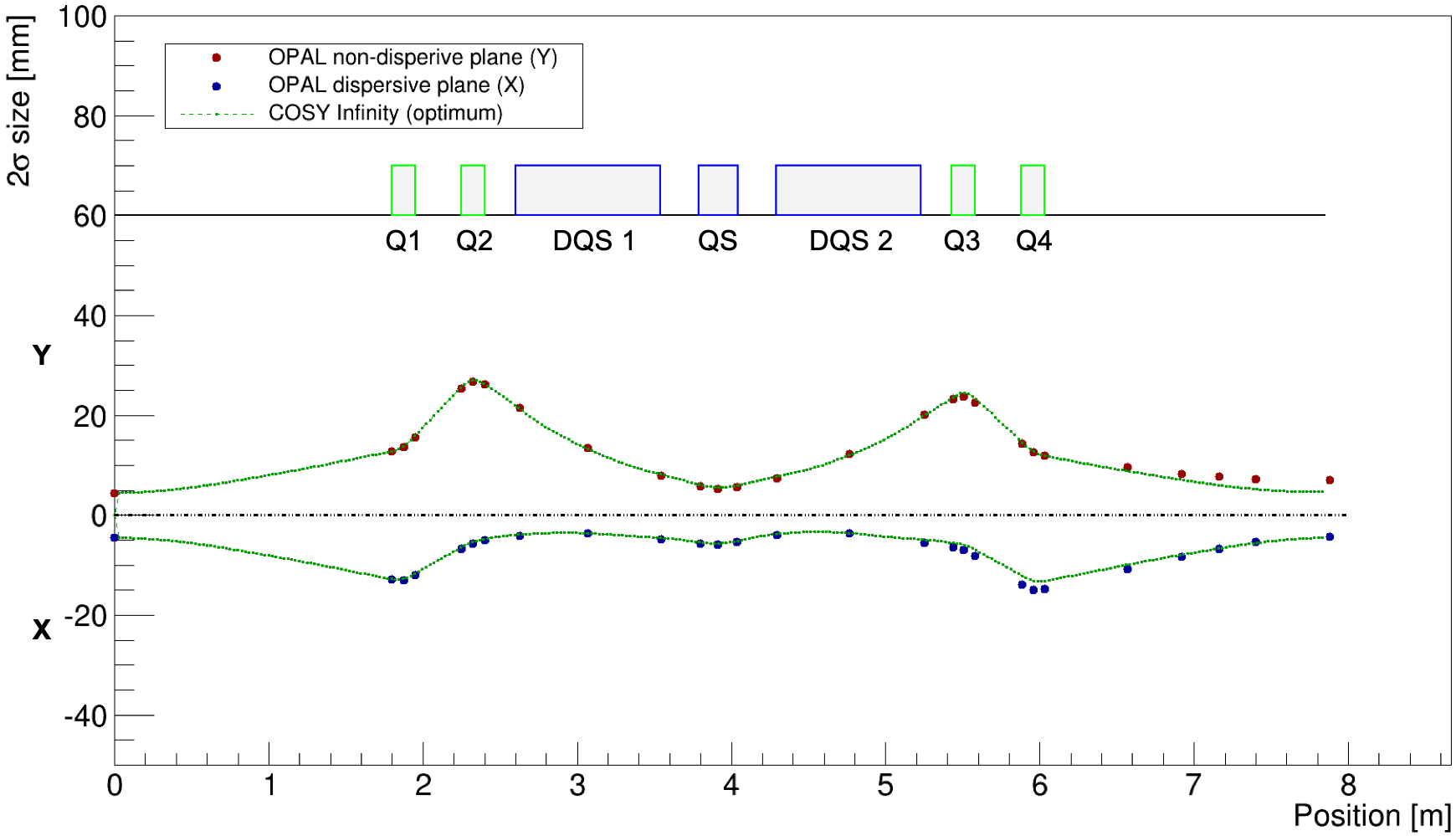} 
\caption{Comparison of \textit{OPAL} simulation and the beam envelope corresponding to the model optimized in~\textit{COSY Infinity} for the nominal momentum. Negative values correspond to the dispersive (X) plane, while positive values correspond to the non-dispersive (Y) plane. The lattice structure is highlighted.}
\label{fig:opal_envelope}
\end{figure}

The last step of our design, as depicted in figure~\ref{fig:design}, is particle tracking in~\textit{OPAL}. The aim is to validate the solution and apply corrections if necessary. A 3D magnetic field map from~\textit{Opera} is used to track particles. Around the nominal momentum corresponding to the beam energy of 185~MeV, a Gaussian beam distribution is produced. The phase space parameters at the starting point (the location of the beam size collimator) are the same as for the~\textit{COSY Infinity} simulation. Since~\textit{OPAL} simulations are based on time integration, the beam size along the lattice is evaluated by means of geometrical probes which record particle hits independently of their time of flight. The probes are defined perpendicular to the reference track at several locations. The 2$\sigma$ beam size at each probe is evaluated from the corresponding distribution of particle hits. In figure~\ref{fig:opal_envelope} the obtained results are compared with the beam envelope for the optimized segmented model describing magnets and fringe fields from~\textit{COSY Infinity}. 
Although a further fine tuning of the solution is needed, it can be concluded that a good agreement has been achieved already. According to the~\textit{OPAL} simulation, the 2$\sigma$ beam sizes at the isocenter are 4.5~mm and 7.0~mm in the X and Y planes, respectively. While in the X plane it agrees with the optimum solution from~\textit{COSY Infinity}, the beam size in the Y plane should be decreased by 30\%. This, however, could be relatively easily achieved by iterative tuning of the lattice parameters in \textit{Opera} following \textit{OPAL} simulation outcomes.

\begin{figure}
\centering
\includegraphics[width=0.8\textwidth]{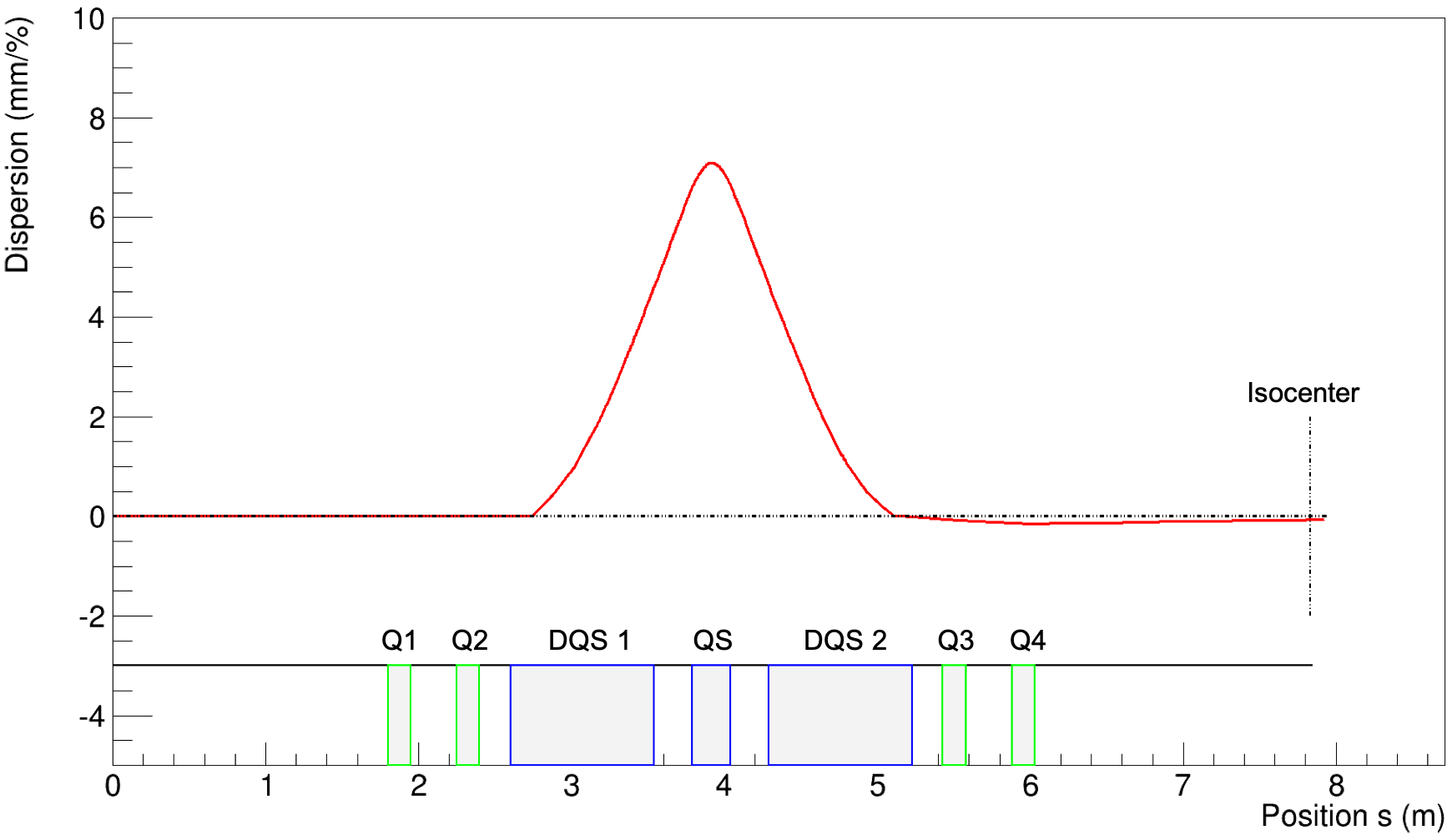} 
\caption{Dispersion as a function of the position along the reference track. The lattice structure is highlighted.}
\label{fig:dispersion}
\end{figure}

Since we aim at a large momentum acceptance, one of the most important parameters to be controlled is dispersion. This is realized by tracking two particles in~\textit{OPAL}. One has the nominal momentum and the second is injected into the lattice with a momentum deviation $dp/p=1\%$. At the starting point both particles are at the magnetic axis and have zero slope $dx/ds$. The off-momentum particle is tracked with a higher granularity so that for each time step of the reference particle a vector perpendicular to its track can be found and the distance between the trajectories of on- and off-momentum particles can be determined. The geometrical dispersion is then plotted as function of the position along the reference track, as shown in figure~\ref{fig:dispersion}. The result indicates that dispersion is correctly suppressed just after DQS 2. The expected symmetry with respect to the center of QS is also visible. The dispersion at the isocenter is as low as 0.06~mm/\%. This could be further improved in the already mentioned final fine tuning.     

\begin{figure}
\centering
\includegraphics[width=0.8\textwidth]{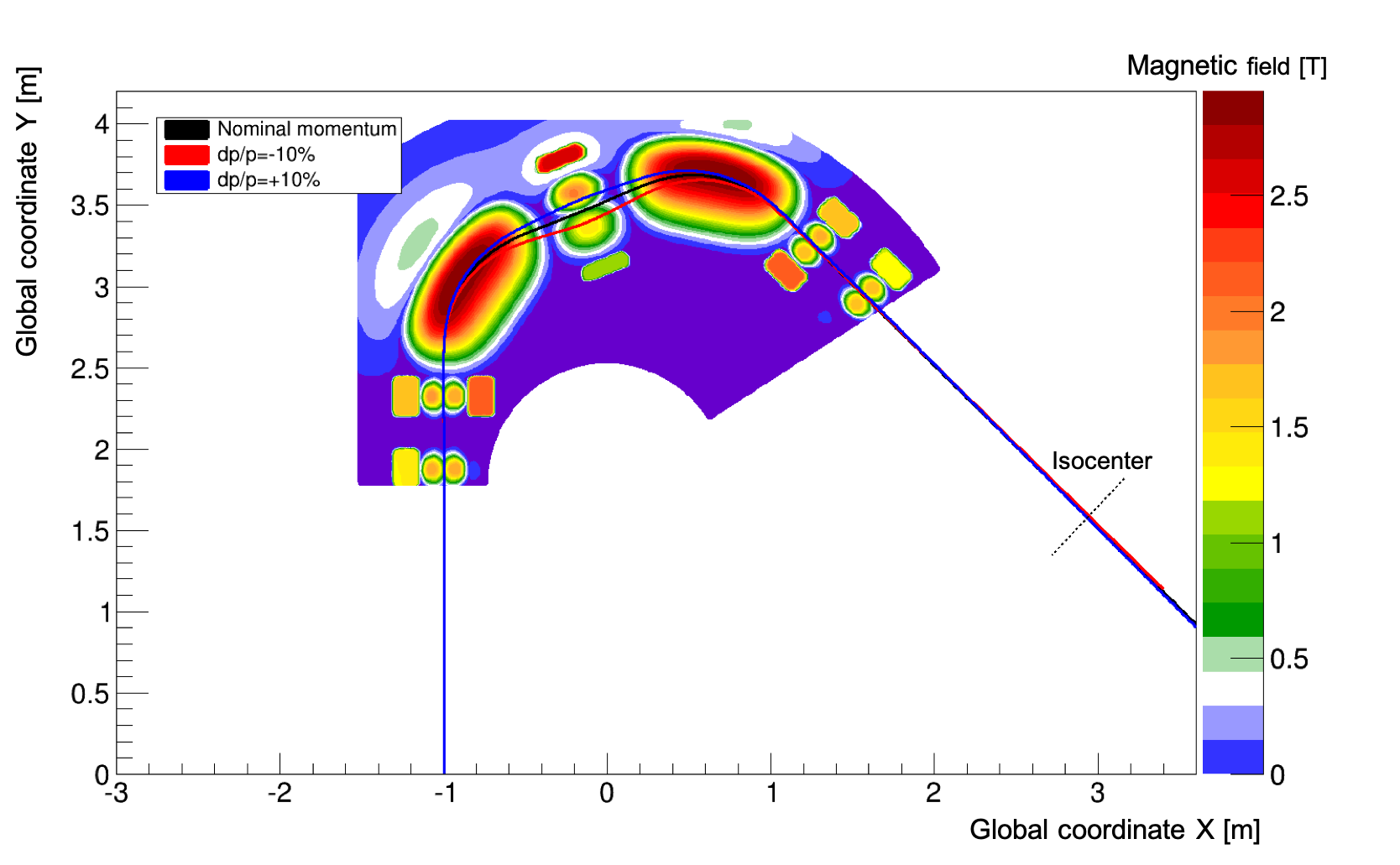} 
\caption{\textit{OPAL} tracking of single particles with different momenta in a 3D magnetic field map.}
\label{fig:opal_global}
\end{figure}

In order to check how particles with larger momentum deviations behave, the tracking of single particles is performed. Figure~\ref{fig:opal_global} shows trajectories of three particles - one with the nominal momentum and two with $dp/p=\pm10\%$. Already at this stage of the design, a good result is achieved. After the fine tuning the beam envelopes for large momentum deviations would have to be verified.

\newpage
\section{Conclusions and outlook}
In this contribution we have presented a design of a gantry for proton therapy in which the last bending section consists of superconducting magnets with a very large momentum acceptance of $\pm15\%$.  Therefore, most of the treatments can be performed without changing the magnetic field to synchronize with energy modulation. Combining with a degrader that can make very fast energy steps, this gantry will be able to perform very rapid energy variations at the patient, reducing the irradiation time.

An important additional advantage of having the degrader mounted on the gantry, is that the footprint of cyclotron driven proton therapy facilities can be reduced substantially by refraining from the usual degrader and following energy selection system.

In this contribution the design process of the magnets and the beam transport has been described. Different codes have been used in this iterative process. The codes~\textit{COSY Infinity} and~\textit{OPAL} have been used to design the optics of the achromatic beam transport and to perform particle tracking in the magnetic fields, which have been calculated from the magnets designed in~\textit{Opera}.

Currently the design has reached a state, which can be transferred to a magnet fabrication design phase. Of course, to optimize the final shaping and dimensioning of the coils and the iron, it will need another iteration step to keep the desired beam optics. Also the gantry mounted fast degrader, the associated collimation system and 2D-lateral scanning system still have to be designed.

Summarizing, the advantages of the SC-gantry design presented here consist of a reduction of weight by up to a factor of 10, a 20\% smaller gantry dimension and an energy acceptance allowing treatments with almost no changes of the magnetic field to follow the energy modulation, so that very fast dose delivery methods can be applied.
\section*{References}
\bibliography{bibliography}

\end{document}